\documentclass{rome_proceedings}

\newcommand{\be}{  \begin{equation}}
\newcommand{\ee}{    \end{equation}}
\newcommand{\bc}{  \begin{center}}
\newcommand{\ec}{    \end{center}}
\newcommand{\bt}{  \begin{tabular}}
\newcommand{\et}{    \end{tabular}}
\newcommand{\bes}{ \begin{equation*}}
\newcommand{\ees}{   \end{equation*}}
\newcommand{\beq}{ \begin{eqnarray}}
\newcommand{\eeq}{   \end{eqnarray}}
\newcommand{\ba }{    \begin{array}}
\newcommand{\ea }{      \end{array}}
\newcommand{\beqs}{\begin{eqnarray*}}
\newcommand{\eeqs}{  \end{eqnarray*}}

\newcommand{\Lc}{\Lambda^{+}_{c}}

\newcommand{\Ds}{\rm \mbox{D}_s}

\newcommand{\Dsp}{\mbox{D}_s^+}

\newcommand{\Dstar}{\mbox{D}^{\ast}}

\newcommand{\Dstarp}{\mbox{D}^{\ast +}}

\newcommand{\Dp}{\mbox{D}^{+}}

\newcommand{\Do}{\mbox{D}^{0}}
\newcommand{\Dob}{\overline{\mbox{D}^{0}}}
\newcommand{\Kp}{\mbox{K}^{+}}
\newcommand{\Km}{\mbox{K}^{-}}

\newcommand{\ddbar}{{\rm D}^{\circ}-\overline{{\rm D}}^{\circ}}

\newcommand{\Zz}{{\rm Z}}

\newcommand{\Vud}{\left | {\rm V}_{ud} \right |}
\newcommand{\Vus}{\left | {\rm V}_{us} \right |}

\begin{document}

\title{Tau and Charm physics highlights}

\author{P. Roudeau}

\address{{\bf Laboratoire de l'Acc\'el\'erateur Lin\'eaire},\\ CNRS-IN2P3 et Univ. de Paris-Sud, B\^at. 200, BP 34 -  91898 Orsay cedex, France
\\E-mail: patrick.roudeau@cern.ch}

\twocolumn[\maketitle\abstract{
In $\tau$ physics, we are at the frontier between the completion of the LEP
program and the start of analyses from $b$-factories, which are expected
to produce results in the coming years. Nice results from CLEO are 
steadily delivered in the meantime. For charm, impressive progress
have been achieved by fixed target experiments in the search
for CP violation and $\ddbar$ oscillations. First results from
$b$-factories demonstrate the power of these facilities in such areas.
The novel measurement of the $\Dstar$ width by CLEO happens to be rather 
different from current expectations. The absence of a charm factory explains 
the lack or the very slow progress in the absolute scale determinations
for charm decays.}]

\section{Tau physics}

$\tau$ physics is an extremely rich area. In the following, I will consider the
tests of lepton couplings universality to weak vector bosons, in light
of recent measurements. 
In a different domain, $\tau$ decays into two pions play a special role
as they can be related to $e^+-e^-$ annihilation into hadrons (I=1 component)
and thus provide an accurate determination of the pion form factor. It can
be noted that, in $\tau$ decays, measurements are obtained within the same experimental running conditions and that hadronic decays are normalized
relative to leptonic decays. These favourable circumstances allow, in general,
a better control of systematics as compared with 
experiments operating at low energy $e^+-e^-$ machines which
need to run at different energies and have to determine an 
absolute normalization
for each energy point. These properties have been used to reduce 
uncertainties related to the virtual photon hadronic component in the 
evaluation of, for instance, the muon anomalous magnetic moment. 
Hadron production
in $\tau$ decays, through the charged weak current,
is theoretically ``clean'' and allows one to measure $\alpha_s(m^2_{\tau})$
and to determine the value of the strange quark mass.

Because of the time allocated for this presentation, I was unable to cover
other topics such as the detailed spectroscopy of hadronic states in 
$n~\pi$ ($n \ge 3$)  decays, recent results on the Lorentz structure in 
leptonic decays \cite{ref:michel} and new measurements
of topological branching fractions \cite{ref:topol}.

Results presented in this review have been obtained by analysing a few $10^5$
$\tau$ pairs in each LEP collaboration and a few $10^6$ events in CLEO.
In future  a few $10^8$ events will be available at $b$-factories.

\subsection{Neutral current universality}\label{subsec:nunivers}
LEP results have been finalized, 
the last analyses from ALEPH have been completed
\cite{ref:alephpolartau}. An important property of the $\tau$ lepton is
that its polarisation can be measured from distributions of
kinematical variables obtained from the four-momenta of 
its decay products. The variation of this polarization versus the
angle ($\theta$) between the $\tau^-$ and the incident electron beam
directions (Eq. (\ref{eq:ptau}))
allows one to extract, separately, the electron (${\cal A}_e$) and $\tau$
(${\cal A}_{\tau}$) asymmetries.
\begin{equation}
P_{\tau}(\cos{\theta})=- \frac{{\cal A}_{\tau}(1+\cos^2{\theta})+{\cal A}_e
(2 \cos{\theta})}
{(1+\cos^2{\theta})+\frac{4}{3}{\cal A}_{FB}(2 \cos{\theta})}
\label{eq:ptau}
\end{equation}
with:
\begin{equation}
{\cal A}_l = \frac{2 g_V^l g_A^l}{(g_V^l)^2+(g_A^l)^2}
\label{eq:aetau}
\end{equation}
$ g_V^l$ and $ g_A^l$ are, respectively, the vector and axial
vector couplings of the lepton $(l)$ to the $\Zz$ boson.

\begin{figure}[h!]
\epsfxsize180pt
\figurebox{180pt}{180pt}{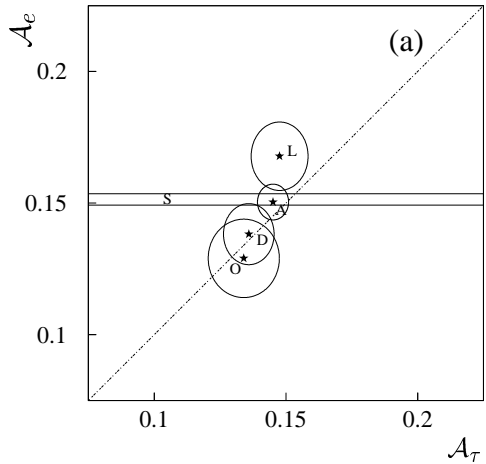}
\caption{
Comparison of the ${\cal A}_e$ and ${\cal A}_{\tau}$
measurements by the ALEPH (A)\protect\cite{ref:alephpolartau}, 
DELPHI (D)\protect\cite{ref:delphipolartau},
L3 (L)\protect\cite{ref:lpolartau}, and OPAL (O)\protect\cite{ref:opalpolartau}
collaborations. The ellipses are one standard error contours (39$\%$ CL).
The horizontal lines represent the SLD (S)\protect\cite{ref:sldpolartau}
${\cal A}_e$ measurement plus 
and minus
one standard deviation. Figure from \protect\cite{ref:alephpolartau}.
\vspace*{-0.2cm}
}
\label{fig:aetau}
\end{figure}

One may note, from figure \ref{fig:aetau}, that all measurements are
compatible and also that ALEPH results have the best accuracy. This comes 
mainly from a larger sample of analysed events (such as for instance
$\tau^- \rightarrow a_1^- \nu_{\tau};~a_1^- \rightarrow \pi^- \pi^0 \pi^0$
decays) and from a better control of systematics. It is apparent also
that the use of polarized beams
from SLD, brings an important constraint on ${\cal A}_e.$

\begin{figure}[h!]
\vspace*{-0.5cm}
\epsfxsize180pt
\figurebox{180pt}{180pt}{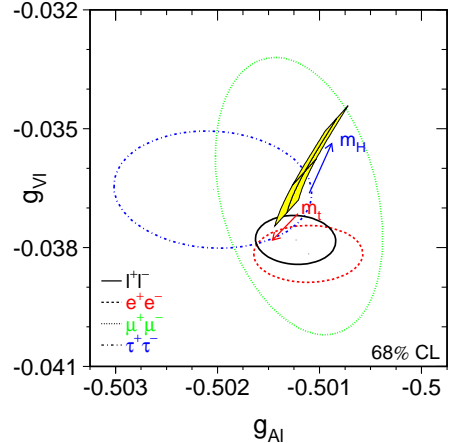}
\caption{Effective lepton couplings. The ellipses are one standard deviation
contours (39$\%$ CL). The shaded area indicates the Standard Model 
expectation. Figure from \protect\cite{ref:lepewwg}.}
\label{fig:gagv}
\end{figure}

The measurements of production rates, forward-backward asymmetries and,
in case of $\tau$ leptons,
of the polarization, can be used to extract values for the vector
and axial-vector couplings of leptons to the $\Zz$ boson.
Figure \ref{fig:gagv} indicates that the three lepton families select 
compatible
regions. In terms of accuracy, the $\tau$ has an intermediate position
between the electron, which benefits from SLD \cite{ref:sldpolartau}
results on initial
state polarization, and the muon for which no information on polarization
is available. The overall average favours a rather light Higgs boson mass
as is usually said.

\subsection{Charged current universality}\label{subsec:cunivers}
Several measurements can be used to study the universality of lepton
couplings to $W$ bosons such as $\tau \rightarrow$ leptons, 
$\tau \rightarrow \pi({\rm K}) \nu_{\tau}$, ... in the following,
among all possible tests,
the most accurate results have only been considered.

$\tau$ leptonic decays have been measured into electron and muon
final states. The present accuracies on the corresponding branching
fractions are dominated by LEP results which are almost in their final
form. 
The universality in charged current couplings can be studied,
considering the parameters $g_{e,\mu,\tau}$ defined as:
\vspace*{0.2cm}
\begin{equation}
{\rm G}_{\tau\ell} = \frac{1}{\sqrt{2}} 
\left ( \frac{g_{\tau} g_{\ell}}{4 m_W^2}\right )= {\rm G}_F
\label{eq:gl}
\vspace*{0.2cm}
\end{equation}
The theoretical expression for the leptonic $\tau$ branching fraction 
is then:
\begin{eqnarray}
{\rm BR}(\tau^-& \rightarrow& \ell^- \overline{\nu}_{\ell} \nu_{\tau})=
\frac{{\rm G}^2_{\tau\ell} m_{\tau}^5 \tau_{\tau}}{192 \pi^3}
f \left( \frac{m_l^2}{m_{\tau}^2}\right ) \nonumber \\
&\times&
\left ( 1 + \frac{3 m_{\tau}^2}{5 m_W^2}\right )
(1 + \delta_{QED})
\label{eq:taubr}
\end{eqnarray}
In this expression, $f$, is a phase space correction to account for
the final state charged lepton mass:
\begin{equation}
f(x)=1-8x+8x^3-x^4-12x^2 \ln{x}
\label{eq:fml}
\end{equation}
and $\delta_{QED}$ contains radiative corrections \cite{ref:corrqed}:
\begin{equation}
\delta_{QED}=\frac{\alpha(m_{\tau})}{2 \pi}
\left ( \frac{25}{4}-\pi^2 \right )
+6.743 \left ( \frac{\alpha(m_{\tau})}{\pi}\right )^2.
\label{eq:deltaqed}
\end{equation}
From present averages \cite{ref:avgbrltau} of existing measurements:
\begin{equation}
{\rm BR}(\tau^- \rightarrow e^- \overline{\nu}_{e} \nu_{\tau})=(17.804 \pm 0.051)\%
\label{eq:brtaue}
\end{equation}
\begin{equation}
{\rm BR}(\tau^- \rightarrow \mu^- \overline{\nu}_{\mu} \nu_{\tau})=(17.336 \pm 0.051)\%
\label{eq:brtaumu}
\end{equation}
and taking into account the expected difference coming from final lepton masses
one obtains:
\begin{equation}
\frac{g_{\mu}}{g_e}= 1.0006 \pm 0.0021
\label{eq:gmuovge}
\end{equation}
which can be compared with the value of $ 1.0023 \pm 0.0016$ deduced
from measurements of $\pi \rightarrow \ell \overline{\nu}_{\ell}$ decays
and
using the determination of radiative corrections given in \cite{ref:pinu}.

\begin{table*}[t]
\begin{eqnarray}
\frac{{\rm d}\Gamma(\tau^- \rightarrow \pi^- \pi^0 \nu_{\tau})}{{\rm d}q^2}
& =&
\frac{G_F^2 \Vud^2 S_{EW}^{\pi\pi}}{32 \pi^2 m_{\tau}^3}
\left ( m_{\tau}^2-q^2 \right )^2 \left ( m_{\tau}^2 +2 q^2 \right )
v^{\pi\pi^0}(q^2) \nonumber \\
\sigma(e^+e^- \rightarrow \pi^+ \pi^-) & =&
\left ( \frac{4 \pi^2 \alpha^2_{em}}{s}\right ) v^{\pi\pi}(s)
\label{eq:taupipio}
\end{eqnarray}
\end{table*}

\begin{figure}
\epsfxsize180pt
\figurebox{180pt}{180pt}{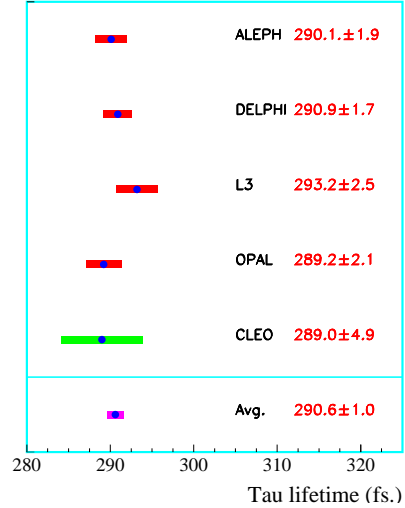}
\caption{Present measurements of the $\tau$ lifetime and global average
(units are $fs$).}
\label{fig:taulife}
\end{figure}

$\tau$ lifetime measurements at
LEP are also almost final, the
only remaining unpublished result, from DELPHI
\cite{ref:delphitau}, has been delivered at this conference. 
The present accuracy on the $\tau$ lifetime is dominated
by LEP experiments as illustrated in figure \ref{fig:taulife}. It must
be noted that the published CLEO result is from 1996, an updated analysis will have an improved accuracy.
The equality of $\tau$ and $\mu$ couplings can be verified by comparing 
the value of the $\tau$ electronic branching fraction obtained within
this hypothesis and the measured one:
\begin{eqnarray}
{\rm BR}_e^{th.} & =&
\left ( \frac{m_{\tau}}{m_{\mu}}\right )^5 \frac{\tau_{\tau}}{\tau_{\mu}}
\times 1.0005 \nonumber \\
 &=& 0.17807 \pm 0.00061(\tau_{\tau})\nonumber \\
&&   ~~~~~~~\pm 0.00015(m_{\tau}) \nonumber \\
\label{eq:brtaueth}
\end{eqnarray}
${\rm BR}_e^{th.}$ has been obtained by applying Eq. (\ref{eq:taubr})
to $\tau^- \rightarrow e^- \overline{\nu}_e \nu_{\tau}$ and
$\mu^- \rightarrow e^- \overline{\nu}_e \nu_{\mu}$, assuming
$g_{\tau}=g_{\mu}$.
\begin{equation}
 {\rm BR}^{exp.}_e =
0.17814 \pm 0.00036
\label{eq:brtaueex}
\end{equation}
These two values are in agreement and, because of the accurate determination
of the $\tau$ lepton mass by BES \cite{ref:bestau}, 
the main uncertainty, in this comparison,
is given by $\tau$ lifetime measurements.
The experimental result, quoted in Eq. (\ref{eq:brtaueex}),
 is the average of the two values given in 
Eq. (\ref{eq:brtaue}) and (\ref{eq:brtaumu}) and has been obtained assuming
$e$-$\mu$ universality.

\subsection{The $\tau^- \rightarrow \pi^- \pi^0 \nu_{\tau}$ channel} 
\label{subs:taupipi}
This decay mode of the $\tau$ lepton is of peculiar interest as it can
be related to the two-pion channel produced in $e^+-e^-$ annihilation,
corresponding to isospin one final states, through the
Conserved Vecteur Current hypothesis. 

The C.V.C. hypothesis corresponds to the following equality between the
spectral functions:
\begin{equation}
v^{\pi\pi}_{I=1}(q^2)=v^{\pi\pi^0}(q^2)~(CVC).
\end{equation}
Spectral functions contain the decay dynamics and are defined
in Eq. (\ref{eq:taupipio}).

\begin{figure}[h]
\epsfxsize160pt
\vspace*{-0.2cm}
\hspace*{-0.5cm}
\rotatebox{-90}{\figurebox{180pt}{180pt}{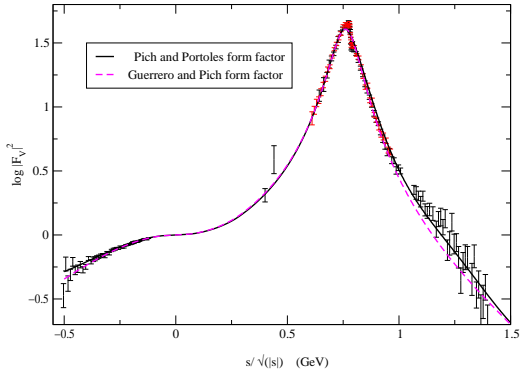}}
\vspace*{0.3cm}
\caption{Comparison of the result of the fit explained in \protect\cite{ref:portoles}
with the experimental data on $F_{\pi}(s)$ from
$e^+ e^- \rightarrow \pi^+ \pi^-$ (time-like) \protect\cite{ref:eepipi}
and $e^- \pi^{\pm} \rightarrow e^- \pi^{\pm}$ (space-like) \protect\cite{ref:epi}.
The result of the fit which has three parameters is compared with
the parametrization \protect\cite{ref:guerre} which depends only on the value
of the $\rho$ mass. In the region below 0.8 GeV both curves are 
indistinguishable. Figure from \protect\cite{ref:portoles}.}
\label{fig:portoles}
\end{figure}

The electromagnetic pion form factor can be, in turn, related to
these distributions through the equality:
\begin{equation}
v^{\pi\pi}(q^2)=\frac{1}{12 \pi} \left | F_{\pi}(q^2)\right |^2
\left ( \frac{2 p_{\pi}}{\sqrt{q^2}}  \right )^3,~q^2=m^2_{\pi\pi}
\end{equation}
It is normalized such that $F_{\pi}(0)=1$. A recent model
independent parametrization of the pion form factor, based on unitarity, 
analyticity and on the chiral behaviour of QCD at low energy,
has been proposed \cite{ref:portoles}
which depends only on the $\rho$ mass value. It is well in agreement
with the measurements up to $q^2 \sim 1$ GeV$^2$ as shown in figure
\ref{fig:portoles}. Such comparisons can be found also in the 
presentation of J. Bijnens at this Conference \cite{ref:bijnensa}.

\begin{figure}[h]
\epsfxsize180pt
\figurebox{180pt}{180pt}{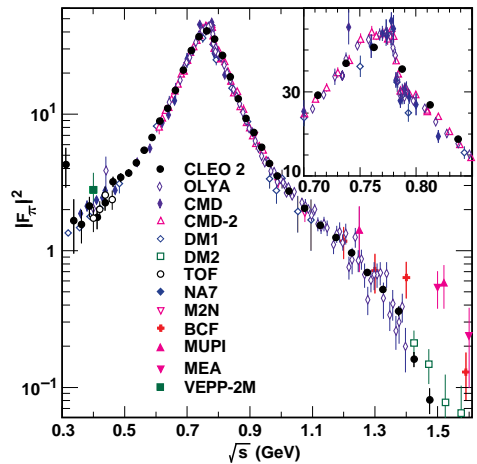}
\vspace*{0.2cm}
\caption{Comparison of $\left | F_{\pi} \right |^2$ as determined
from CLEO data (filled circles), with that obtained from 
$e^+ e^- \rightarrow \pi^+ \pi^-$ cross sections (other symbols). 
The inset is a blow-up of the region near the $\rho$ peak, where 
$\rho-\omega$ interference is evident in the $e^+~e^-$ data.
Figure from \protect\cite{ref:cleotaupipio}.}
\label{fig:cleotaupipio}
\end{figure}

The most recent measurement, obtained by CLEO 
\cite{ref:cleotaupipio}, is shown in figure \ref{fig:cleotaupipio}.

The C.V.C. hypothesis is expected to be violated at an extremely low
level: ${\cal O}(m_u-m_d)^2$. Experimentally, measurements obtained in
$\tau$ decays and in $e^+-e^-$ annihilation have to be corrected
for I-spin violating effects coming from electromagnetic interactions as:
$\omega^0 \rightarrow \pi^+ \pi^-$, $m_{\pi^+} \neq m_{\pi^0}$
and $(m,\Gamma)_{\rho^-} \neq (m,\Gamma)_{\rho^0}$.
After these corrections, the difference between the measured \cite{ref:alephpp}
$\tau^- \rightarrow \pi^- \pi^0 \nu_{\tau}$ branching fraction and the value
deduced from $e^+-e^-$ measurements, which includes new results from
CMD-2 \cite{ref:mdtwo}, is evaluated to be $(0.37 \pm 0.29)\%$ 
\cite{ref:eidelman}.
It may be noted that the fitted mass value of the $\rho$ meson 
$(775 \pm 1)$ MeV/c$^2$ is rather different from the present value
quoted in the main section of PDG2000 \cite{ref:pdg00}
which is $(769.3 \pm 0.8)$ MeV/c$^2$. This last  value is dominated by results
obtained in other experimental conditions than $\tau$ decays or $e^+-e^-$ 
annihilation.
\vspace*{0.2cm}

\subsection{$\alpha_s(m_{\tau}^2)$ measurement}
\label{subs:alphasfrt}
Studies of hadronic $\tau$ decays have been the basis of the most accurate
measurement of the strong coupling constant.

First results on $\alpha_s(m_{\tau}^2)$ were obtained in 1993 from pioneer
analyses done by ALEPH \cite{ref:aleph93} which were based on 
theoretical developments \cite{ref:alphasth} derived in the framework
of the Operator Product Expansion formalism \cite{ref:ope}.
The hadronic $\tau$ decay width, normalized to the leptonic width, $R_{\tau}$,
can be expressed, using the O.P.E. formalism in terms
of a contour integral at $s \simeq m^2_{\tau}$.
\begin{eqnarray}
R_{\tau}&=&3\left ( \Vud^2 + \Vus^2 \right ) S_{EW}\nonumber \\
 && \times \left [ 1 +\delta^{\prime}_{EW} + \delta^P +\delta^{NP}  \right ]
\label{eq:rtauth}
\end{eqnarray}
with: $ S_{EW}=1.0194 ~{\rm and}~\delta^{\prime}_{EW}=0.0010$.
$S_{EW}=1+2 \left ( \frac{\alpha}{\pi}\right )\ln{\frac{m_Z}{m_{\tau}}}$
represents a short-distance correction due to virtual particles with energies
ranging from $m_{\tau}$ to $m_Z$. Summing all leading logarithms of
$m_Z/m_{\tau}$ gives the quoted value \cite{ref:marcia}.
The remaining radiative correction, 
$\delta^{\prime}_{EW}=\frac{5}{12}\frac{\alpha(m_{\tau})}{\pi}$,
can be found in \cite{ref:braaten}.
From theory, it is expected that the dominant corrections to the naive
expectation $R_{\tau}={\rm N}_c=3$ originates from perturbative QCD:
\begin{eqnarray}
\delta^P &= & \frac{\alpha_s(m_{\tau}^2)}{\pi}
+5.2023 \left(\frac{\alpha_s(m_{\tau}^2)}{\pi}  \right )^2 \nonumber \\
&+& 26.366 \left(\frac{\alpha_s(m_{\tau}^2)}{\pi}  \right )^3 \nonumber \\
&+& (78.003+K_4) \left(\frac{\alpha_s(m_{\tau}^2)}{\pi}  \right )^4\nonumber \\
 &+&{\cal O}\left(\frac{\alpha_s(m_{\tau}^2)}{\pi}  \right )^5
\label{eq:ratauth}
\end{eqnarray}
$K_4$ is unknown, different estimates give a value of the order of $30$,
in the evaluation of systematic uncertainties, it has been assumed that
$K_4$ varies within the range: $50\pm50$.
Corrections of non-perturbative origin are expected to be suppressed
as $m_{\tau}^{-n}$ with $n \ge 2$:
\begin{equation}
\delta^{NP} \simeq {\cal O}\left(\frac{m_q}{m_{\tau}}  \right)^2
+ \sum_{D=4,6,.} C_D(\mu) \frac{<O>_D}{m_{\tau}^D}
\label{eq:nonp}
\end{equation}
where $C_D(\mu)$ are short distance coefficients given by theory.
The first term in this expression comes mainly from the 
value of the strange quark mass because effects from lighter quarks can be
neglected. This property will be exploited in Sec. \ref{sec:ms}.
Experimentally it is rather easy to measure $R_{\tau}$ as it depends only
on the $\tau$ leptonic branching fraction:
\vspace*{0.2cm}
\begin{eqnarray}
R_{\tau}&=&\frac{\Gamma(\tau \rightarrow X_h \nu_{\tau})}
{\Gamma(\tau^- \rightarrow e^- \overline{\nu}_{e} \nu_{\tau})} =
\frac{1}{B_e}-1-\frac{B_{\mu}}{B_e} \nonumber \\
& = & 3.641 \pm 0.011.
\vspace*{0.2cm}
\end{eqnarray}

 Main progress during the past few years have consisted in measuring
the dominant contributions from non-perturbative corrections
\cite{ref:nonpert} \cite{ref:alephaa}
demonstrating that they are under control.
In this purpose, the three independent contributions to the hadronic
final state, in $\tau$ decays, have been distinguished. The strange
component, originating from Cabibbo suppressed $W^- \rightarrow \overline{u}s$
decays, has been isolated by considering events
with an odd number of kaons. In this way, non-perturbative QCD corrections
of order ${\cal O}(1/m_{\tau}^2)$ are eliminated,
once corresponding events are removed from the analysis.
The vector and axial-vector
hadronic components are then separated by considering events with, 
respectively, an even or an odd number of pions. Corrections have to be applied
to account for isospin violating decays of $\eta$ and $\omega$ mesons
and to distribute events with kaons \cite{ref:rouge} over the two components. 
In a way, which is similar to the two pion
final state, spectral functions can be defined which contain all the
decay dynamics (see Eq. (\ref{eq:voua})).

\begin{figure}[h]
\epsfxsize180pt
\figurebox{180pt}{180pt}{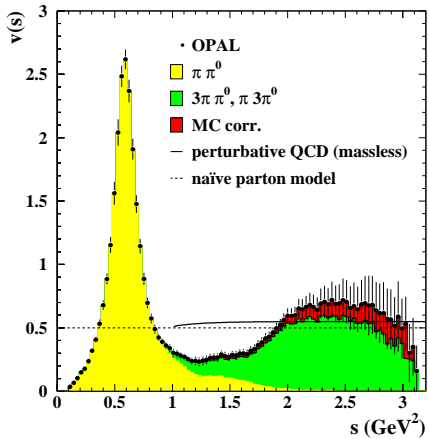}
\caption{The vector spectral function measured by OPAL.
Shown are the sums of all contributing channels as data points. 
Some exclusive contributions are shown as shaded areas. The naive
parton model prediction is shown as dashed line, while
the solid line depicts the perturbative, massless QCD prediction.
Figure from \protect\cite{ref:opalalphas}.}
\label{fig:vopal}
\end{figure}

\begin{figure}[h!]
\epsfxsize180pt
\figurebox{180pt}{180pt}{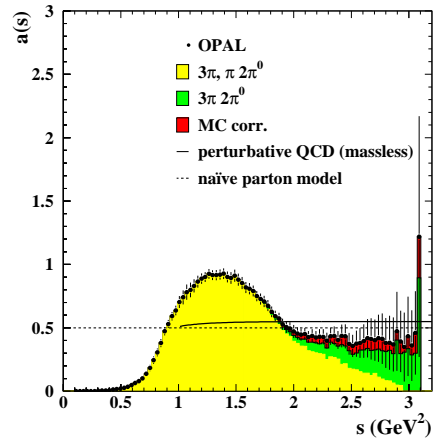}
\caption{The axial-vector spectral function measured by OPAL.
The same conventions have been used as in figure \ref{fig:vopal}.
The pion pole has been subtracted. Figure from \protect\cite{ref:opalalphas}.}
\label{fig:aopal}
\end{figure}

\begin{figure}
\epsfxsize180pt
\figurebox{180pt}{180pt}{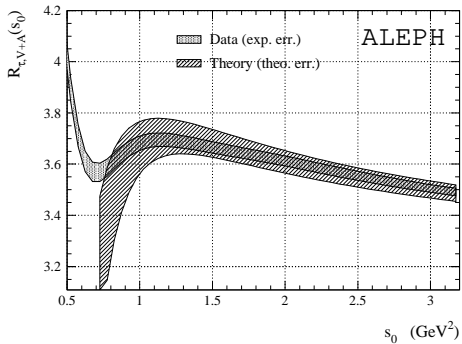}
\caption{The ratio $R_{V+A}$ versus the square of the ``light-$\tau$''
mass $s_0$. The curves are plotted as error bands to emphasize their strong
point-to-point correlations in $s_0$. Also shown is the theoretical 
prediction. Figure from \protect\cite{ref:alephaa}.}
\label{fig:stabvpa}
\end{figure}

\begin{table*}[tb!]
\begin{equation}
v_1(s)/a_1(s) = \frac{m_{\tau}^2}{6 \Vud^2 S_{EW}}
\frac{1}{\left( 1-\frac{s}{m_{\tau}^2}\right)^2 
\left (1 + \frac{2 s}{m_{\tau}^2}  \right )}
\frac{{\rm BR}(\tau^- \rightarrow V^-/A^- \nu_{\tau})}
{{\rm BR}(\tau^- \rightarrow e^- \overline{\nu}_{e} \nu_{\tau})}
\frac{1}{{\rm N}_{V/A}} \frac{{\rm dN}_{V/A}}{{\rm d}s}
\label{eq:voua}
\end{equation}
\end{table*}
Recent results, presented by OPAL \cite{ref:opalalphas} 
are shown in figures \ref{fig:vopal}
and \ref{fig:aopal}.
The dominant contributions
of the $\rho$ and $a_1$ mesons are apparent, at low masses, respectively in
the $v_1(s)$ and $a_1(s)$ spectral functions. It can be noted also that,
at large mass values, the two functions are approaching the expectations from QCD.
Differences relative to perturbative QCD evaluations are expected to be
of non-perturbative origin and their contributions to 
$R_{\tau}$ have been parametrized in Eq. (\ref{eq:nonp}) using the O.P.E.
formalism. Contributions from the different terms can be enhanced
by considering moments of the spectral functions giving a larger
weight to different regions of the mass distribution.
\begin{table*}[tb!]
\begin{equation}
R^{kl}_{V/A}=\int_0^{s_0\leq m_{\tau}^2} {\rm d}s 
\left( 1-\frac{s}{m_{\tau}^2} \right )^k
\left( \frac{s}{m_{\tau}^2} \right )^l 
 \frac{{\rm d}R_{V/A}}{{\rm d}s}
\label{eq:moments}
\end{equation}
\end{table*}
In this way, coefficients $C_D~(D=4,~6,8)$ have been measured. 
The validity of this approach can also be verified 
\cite{ref:alephstab} by considering
``light-$\tau$'' decays. Instead of evaluating the contour integral,
which corresponds to expression (\ref{eq:rtauth}), at $s=m_{\tau}^2$,
the procedure is repeated for fixed values of $s$ which are smaller than the 
$\tau$ mass and keeping only the parts of the spectral functions which
satisfy this bound. Figure \ref{fig:stabvpa} illustrates the stability
of the method down to hypothetic $\tau$ masses of the order of 1 GeV/c$^2$
when applied to the V+A spectral function. For individual V or A
spectral functions the stability is observed over a lower range,
down to 1.5 GeV/$c^2$.

\begin{table*}[tb!]
\begin{center}
\begin{tabular}{|c|c|c|}
\hline
CLEO \cite{ref:cleotauk}& ALEPH\cite{ref:alephtauk} & OPAL\cite{ref:opaltauk}\\
$(0.346 \pm 0.061)\%$ &$(0.214 \pm 0.047)\%$ & $(0.360 \pm 0.095)\%$ \\
\hline
\end{tabular}
\caption{Measured values for the 
$\tau^- \rightarrow \Km \pi^+ \pi^- \nu_{\tau}$
branching fraction.}
\label{tab:brkpipi}
\end{center}
\end{table*}

The V+A combination appears to be peculiarly stable because
of cancellations between the different non-perturbative contributions.
From this analysis \cite{ref:alephaa},
the following values have been obtained for the
strong coupling constant, extrapolated from the $\tau$ up to the $\Zz$
mass scales,
and for the total contribution of non-perturbative effects:
\begin{eqnarray}
\alpha_s(m_Z^2)&=&0.1202 \pm 0.0008_{exp}\nonumber \\
 &\pm&0.0024_{th}\pm0.0010_{extr}
\end{eqnarray}
\begin{equation}
\delta^{NP}_{V+A}=-0.003 \pm 0.005
\end{equation}
similar results have been obtained by CLEO \cite{ref:nonpert}
and OPAL\cite{ref:opalalphas} collaborations.
\subsection{Strange quark mass measurement}
\label{sec:ms}
The analyses of $\tau$ spectral functions, described briefly in the previous
section, can be sensitive to the value of the strange quark mass through
non-perturbative corrections of order $(m_s/m_{\tau})^2$.
This sensitivity can be enhanced by isolating $\tau$ decays with an 
odd number of kaons. Several branching fractions of such channels have been
measured mainly by ALEPH \cite{ref:alephtauk},
CLEO \cite{ref:cleotauk} and OPAL \cite{ref:opaltauk} 
collaborations. One may note that
there is not a perfect agreement on the values obtained for the easiest
channel ($\tau^- \rightarrow \Km \pi^+ \pi^- \nu_{\tau}$) by these three
experiments (Table \ref{tab:brkpipi}).

\begin{figure}[h]
\epsfxsize180pt
\figurebox{180pt}{180pt}{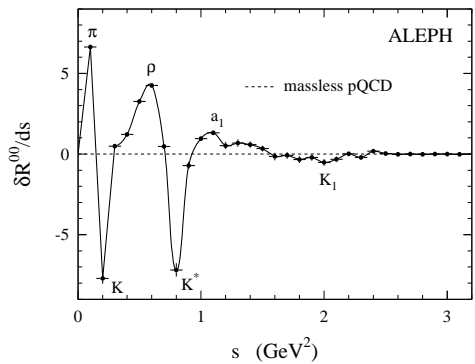}
\vspace*{-0.2cm}
\caption{Value of Eq. (\ref{eq:momms}) corresponding to the difference
between Cabibbo-corrected nonstrange and strange invariant mass spectra. 
To guide the eye, the solid line interpolates between the bins of constant 
0.1 GeV$^2$ width. Figure from \protect\cite{ref:alephtaus}.}
\label{fig:oomomms}
\end{figure}

ALEPH \cite{ref:alephtaus} has also measured the spectral function in 
$\tau^- \rightarrow \overline{u} s \nu_{\tau}$ decays. To reduce the 
effects from perturbative QCD corrections, the following difference
between moments is used:
\begin{equation}
\delta R^{kl}=\frac{R^{kl}_{V+A}}{\Vud^2}
-\frac{R^{kl}_{S}}{\Vus^2} 
\label{eq:momms}
\end{equation}
which is zero in the SU(3) limit.
Figure \ref{fig:oomomms} shows the hadronic mass dependence of 
$\delta R^{00}$. 

\begin{table*}[tb!]
\begin{equation}
a_{\mu}^{had}=\frac{\alpha^2(0)}{3 \pi^3}\int_{4 m^2_{\pi}}^{\infty}
{\rm d}s \frac{K(s)}{s}R(s);
~R(s)=\frac{\sigma(e^+e^- \rightarrow had.)}
{\sigma(e^+e^- \rightarrow \mu^+ \mu^-)}
\label{eq:amur}
\end{equation}
\end{table*}
The main difficulty, in this analysis, comes from the poor convergence of the
perturbative QCD series which depends on the strange quark mass. This 
convergence depends on the choice of the values for $k$ and $l$ indices
and 
a better convergence can be obtained
at the expense of a lower experimental sensitivity 
to $m_s$. The origin of the convergence problem has been explained
in \cite{ref:maltman}.
One can quote the two values from recent results 
\cite{ref:davierms} \cite{ref:kornerms}, obtained for 
different choices of the $k$ and $l$ indices and which appear to be quite compatible:
\begin{eqnarray}
m_s(m_{\tau}^2)& =&(120\pm11_{exp}\pm8_{\Vus}\pm 19_{th})\nonumber \\
{\rm or} & & \nonumber \\
 m_s(m_{\tau}^2) &=&(130\pm27_{exp}\pm 9_{th})
\end{eqnarray}
(values are given in MeV/c$^2$ units).

This is an important result which improves the determination of 
the strange quark mass value and also of values for lighter
quark masses once the ratio $m_s/(m_u+m_d)$ has been taken from
analyses based on chiral perturbation theory \cite{ref:msovermu}.

With larger statistics of Cabibbo suppressed $\tau$ decays, more
accurate determinations of the strange quark mass can be obtained in future.

\subsection{$\tau$ decays and $g-2$}

The anomalous muon magnetic moment $a_{\mu}=\frac{g_{\mu}-2}{2}$
has been recently measured \cite{ref:muexpt}
with improved accuracy and a further reduction of
experimental uncertainties is still expected. To find evidence for
new physics, this measurement has to be compared with theoretical expectations.
The evaluation of $a_{\mu}$ receives three contributions, respectively,
from pure quantum electrodynamics corrections,
from the photon hadronic component and from $W,~\Zz~{\rm or}~H$ exchange:
\begin{eqnarray}
a_{\mu}&=&\left( a_{\mu}^{QED}(\pm 3)+a_{\mu}^{had}(\pm 150)+a_{\mu}^{weak}(\pm 4)\right ) \nonumber \\
 & &\times 10^{-11}.
\label{eq:amuthd}
\end{eqnarray}
The main uncertainty comes from the second component which, itself, can be
splitted into two contributions:
one being due to the photon hadronic vacuum polarization and the other 
which is usually named hadronic light-by-light scattering has to be 
determined from theory \cite{ref:bijnensa}. 
The former contribution is obtained by relating 
its value to the $e^+-e^-$ hadronic annihilation cross section at low energy
(see Eq. (\ref{eq:amur})).

The expression for $K(s)$ can be found for instance in \cite{ref:eidj}
and it appears that 90$\%$ of the integral (\ref{eq:amur})
corresponds to values
of the energy below 1.8 GeV. The uncertainty quoted in Eq. (\ref{eq:amuthd})
corresponds to the analysis \cite{ref:eidj} which is based
on results from $e^+-e^-$ annihilation at low energies, prior to 1995.
Measurements of $\tau$ decays, explained in Sec. \ref{subs:taupipi},
can be used to complement results from $e^+-e^-$ machines, at low energy.
In addition, as $e^+-e^-$ data, available at that time, were incomplete
and suffer from systematics, properties of QCD such as unitarity and analyticity
can be used in the framework of the O.P.E. to apply constraints. 
This approach \cite{ref:davhoun} is rather
similar to the one followed in Sec. \ref{subs:alphasfrt}. For a value
of the energy higher than a typical hadronic scale, taken to be
equal to $1.8$ GeV, the ratio $R(s_0)$ can be expressed
as a contour integral of analytical expressions obtained from
perturbative QCD and from a parametrization on Non-Perturbative contributions
in terms of series in $s_0^{-n}$:
\vspace*{0.2cm}
\begin{equation}
R(s_0)=\frac{1}{2 i \pi}\oint_{|s|=s_0} \frac{{\rm d} s}{s} D(s).
\end{equation}
$D(s)$ is the Adler D-function \cite{ref:adler}. 
Moments can be also defined, in a way similar as in Eq. (\ref{eq:moments})
and mean values of the operators corresponding to
the dominant non-perturbative components can be obtained.
In this approach \cite{ref:davhoun},
the uncertainty on $a_{\mu}^{had}$ is reduced by a factor two
with respect to Eq. (\ref{eq:amuthd}), becoming
equal to $\pm 75 \times 10^{-11}$.
\begin{table*}[tb!]
\begin{equation}
\int_{4 m_{\pi}^2}^{s_0} {\rm d}s f(s) \Im \Pi=
\underbrace{\int_{4 m_{\pi}^2}^{s_0} {\rm d}s [f(s) -P_n(s)]\Im \Pi}_{data}+
\underbrace{\frac{i}{2} \oint_{|s|=s_0} {\rm d} s P_n(s) \Pi^{QCD}}_{theory}
\label{eq:qcdsr}
\end{equation}
\end{table*}

\begin{table*}[tb!]
\begin{center}
\begin{tabular}{|c|c|}
\hline
 Energy (GeV)   & $a_{\mu}^{had} \times 10^{11}$\\
\hline
$(2 m_{\pi}-1.8)_{uds}$ & $6343 \pm 56_{exp} \pm 21_{theo}$   \\
$(1.8 -3.7)_{udsc}$ & $338.7 \pm 4.6_{theo}$   \\
$\psi(1S,2S,3770)+(3.7-5.0)_{udsc}$ & $143.1 \pm 5.0_{exp} \pm 2.1_{theo}$   \\
$(5.0-9.3)_{udsc}$ & $68.7 \pm 1.1_{theo}$   \\
$(9.3-12)_{udscb}$ & $12.1 \pm 0.5_{theo}$   \\
$(12 -\infty)_{udscb}$ & $18 \pm 0.1_{theo}$   \\
\hline
All & $6924 \pm 56 \pm 26$\\
\hline
\end{tabular}
\caption{Contributions to $a_{\mu}^{had}$ from the different energy regions.
The subscripts in the first column give the quark flavours involved
in the calculation. Table taken from \protect\cite{ref:davhounb}.}
\label{tab:differ}
\end{center}
\end{table*}

This procedure is not applied in the region of resonances, below
1.8 GeV and in the vicinity of charm or beauty thresholds.
Further constraints from QCD can be used in these regions, 
originating from QCD sum rules. This method is explained in 
\cite{ref:groko}. It consists in minimizing uncertainties
from experimental results, at the expense of introducing theoretical
uncertainties. Polynomial expressions are determined that mimic the 
singular weight function $K(s)/s$ in Eq. (\ref{eq:amur}). This polynomial
function is subtracted to decrease the importance of the 
dispersion integral which is evaluated from experimental measurements.
In order to compensate for this subtraction (see Eq. (\ref{eq:qcdsr}))
the same polynomial function
is added again, but now, being analytic, its contribution is evaluated
on a circular contour in the complex plane, using a theoretical expression
for the other part of the quantity to integrate.

Contributions to $a_{\mu}^{had}$ from different energy regions of the
$e^+-e^-$ hadronic annihilation cross section,
obtained in the analysis of \cite{ref:davhounb}, are given in Table 
\ref{tab:differ}. The total uncertainty on $a_{\mu}^{had}$
becomes equal to $\pm 62 \times 10^{-11}$ and is dominated by experimental
uncertainties.

In a recent analysis \cite{ref:yndura}, which includes new
and more accurate measurements
from low energy $e^+-e^-$ experiments (CMD-2 \cite{ref:mdtwo} and BES
\cite{ref:besthc}) but in which constraints from QCD sum rules
are not applied,
very similar results have been obtained for the central value and uncertainty
of $a_{\mu}^{had}$.

To improve further in the determination of $a_{\mu}^{had}$,
 additional measurements at low energy, of $e^+-e^-$ 
hadronic annihilation cross sections and $\tau$ decays, with a few
$10^{-3}$ accuracy are needed.

\section{Charm physics}

\begin{table*}[tb!]
\begin{center}
\begin{tabular}{|c|c|c|c|c|}  
\hline
 Experiment & Operation & $\# \Do \rightarrow \Km \pi^+$ & 
$ \sigma_t/\tau_{D^{\circ}}$ & $\sigma(\Delta m)$\\
    & & & & (MeV)\\
\hline
E791  & $\pi^-$ 500 GeV &$\sim 4~ 10^4$ & few $\%$& \\
FOCUS (E831)  & $\gamma$ $\leq$300 GeV & $\sim 10^5$ &$\leq 10\%$ & $\sim 0.7$ \\
SELEX (E781) & $\pi,~\Sigma$ 600 GeV & & $4\%$ & \\
CLEO & $e^+e^-~\sqrt{s}=10.6$ GeV  & $\sim 2-4~10^4$ &$\sim 40\%$ & $\sim 0.2$ \\
BaBar  &$e^+e^-~\sqrt{s}=10.6$ GeV &$\sim 5~ 10^4$ &$\sim 40\%$ &$\sim 0.3$  \\
BELLE  &$e^+e^-~\sqrt{s}=10.6$ GeV & $\sim 9~10^4$ &$\sim 40\%$ & $\sim 0.3$\\
\hline
\end{tabular}
\caption{A few characteristics of experiments contributing in
charm physics.
\label{tab:exptcharm}}
\end{center}
\end{table*}

Ideally, charm physics is an extraordinary place, for testing QCD technologies
\cite{ref:ikario}
as lattice QCD, QCD sum rules or chiral theory because very accurate
measurements can be obtained in charm decays on several key channels.
It can be also considered to extrapolate such results to B physics in which
the measurement of corresponding Cabibbo-Kobayashi-Maskawa matrix elements
requires, usually, a good control of effects from strong interactions.
For instance, at present, the most sensible way to evaluate the $b$-meson 
decay constant
is through the measurement of the $c$-meson decay constant and lattice QCD
\cite{ref:prach}.
The present situation will be summarized in the following and it will be
explained why such a program has not really started in practice.
Another aspect of charm physics is the search for new phenomena. Direct 
searches on $\Do-\Dob$ oscillations and on CP violation in $c$-meson decays
have made impressive progress during the past two years. Another aspect
is related to the possibility of finding new physics effects in other fields,
as $b$-hadron decays, because of the control of hadronic effects gained
from $c$-hadron studies.
 Studies of $c$-hadron decays are also a good place for the 
spectroscopy of hadrons
made of light quarks.
As an example, clean signals from scalar mesons ($\sigma,~\kappa$)
have been observed
in $\Dp$ decays 
\cite{ref:sigma}
but, because of lack of time, I cannot develop these
interesting aspects.

\subsection{Contributing experiments}
A few key parameters illustrating different characteristics of charm
production measured at fixed target experiments and at experiments running
at the $\Upsilon$(4S) are given in Table \ref{tab:exptcharm}.

At present, statistics of reconstructed charm events are rather similar.
This will change during the coming years with the increase of registered
data at $b$-factories. Fixed target experiments had the advantage of measuring
accurately the $c$-hadron decay time. This property is also
crucial, in their analyses, to isolate
clean events samples. Because of the higher relative production rate of
charmed particles, of a better mass resolution and powerful particle 
identification, experiments at the $\Upsilon$(4S) can obtain charm samples
with similar purity even if their decay time resolution is 10 times lower.
The experimental mass resolution is also an important parameter,
as explained in
Sec. \ref{subsec:dstarw}. Only CLEO and, possibly 
in future $b$-factories, seem to
be in a position to measure the $\Dstarp$ width.

\subsection{Charm hadron lifetimes}
\label{susec:clife}
Lifetime measurements are needed to relate experiments, which measure 
branching fractions, and theory, which predicts partial widths.

\begin{figure}[h]
\epsfxsize180pt
\figurebox{180pt}{180pt}{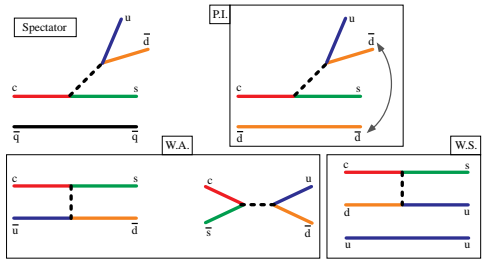}
\caption{Idealized diagrams contributing in $c$-hadron decays.
In the spectator graph, light partons are not contributing in the weak 
process. The Pauli interference (P.I.) graph is active when two
identical partons are present in the final state. The weak annihilation (W.A.)
graphs are helicity suppressed, whereas, for baryons the 
weak scattering (W.S.) is not.}
\label{fig:ddecaydiag}
\end{figure}

Within the O.P.E. formalism, the total decay width
of a charmed hadron can be expressed as a series in $1/m_c$
\cite{ref:bigilife}, the first
contributing correction, related to differences
between mesons and baryons, is of order ${\cal O}(1/m_c^2)$.
\begin{eqnarray}
\Gamma(H_c)&=&\Gamma_{spect.}+{\cal O}(1/m_c^2)\nonumber \\ 
&+&\Gamma_{PI,WA,WS}(H_c)\nonumber \\
&+&{\cal O}(1/m_c^4)
\label{eq:cdecayw}
\end{eqnarray}
The first term in Eq. (\ref{eq:cdecayw}) corresponds to the $c$-quark decay (spectator contribution).
Mechanisms in which the light $c$-hadron constituents 
are involved are of order ${\cal O}(1/m_c^3)$ but can have large
contributions because they are phase-space enhanced (by $16\pi^2$).
These contributions are indicated in figure \ref{fig:ddecaydiag}.
Effects from weak annihilation (W.A.), contributing in $\Do$ and $\Ds$ decays
are expected to be rather small as they are helicity suppressed.
Largest effects are expected from Pauli interference (P.I.) which can be 
destructive and also constructive, in case of baryons, and from 
weak scattering (W.S.) which, for baryons, is no more helicity suppressed.
These considerations are explained in \cite{ref:bigilife}.
\begin{table*}[tb!]
\begin{center}
\
\begin{tabular}{|c|c|}
\hline
lifetime ratio & mechanism \\
\hline
$\tau(\Dp)/\tau(\Do) = 2.54 \pm0.02 $& ${\rm P.I.(-)}$\\ 
$\tau(\Dsp)/\tau(\Do) = 1.206 \pm 0.013 $&$ ?~{\rm interf~W.A./Spect.}$\\ 
$\tau(\Lc)/\tau(\Do) = 0.49 \pm 0.01 $&$ ~{\rm W.S., ~P.I.(-)}$\\ 
$\tau(\Xi_c^+)/\tau(\Lc) = 2.25 \pm 0.14  $&$~{\rm W.S., ~P.I.(\pm)}$\\ 
\hline
\end{tabular}
\caption{Present values for $c$-hadron lifetime ratios.
Expected dominant contributions which can explain deviations from unity
of these ratios are indicated.}
\label{tab:liferat}
\end{center}
\end{table*}

\begin{figure}[h]
\vspace*{-0.5cm}
\epsfxsize180pt
\figurebox{180pt}{180pt}{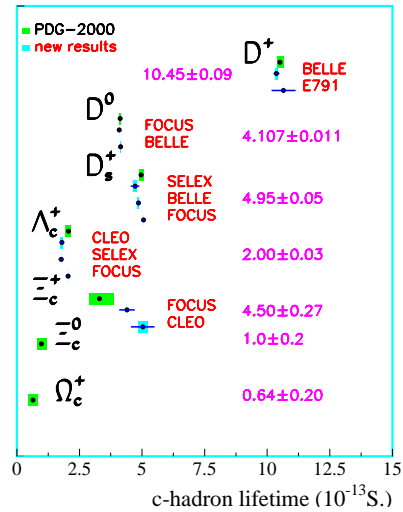}
\caption{New measurements of $c$-hadron lifetimes are compared with PDG2000
averaged values (upper point given for each charmed hadron). For new
results, points with error bars correspond respectively 
to the central values and statistical uncertainties. Shaded zones are for
systematic uncertainties (they correspond to the total uncertainty in PDG 
averages).}
\label{fig:clifemeas}
\end{figure}

At this conference new measurements (see figure \ref{fig:clifemeas})
have been provided by BELLE \cite{ref:belleli}, CLEO \cite{ref:cleoli},
FOCUS \cite{ref:focusli} and SELEX \cite{ref:selexli}
collaborations. Some of these results have a statistical
accuracy similar to or even better than the world average obtained in 
year 2000. Studies of systematic uncertainties are thus crucial and BELLE,
for instance,
has demonstrated that they can be controlled also at a very low level.
An accuracy better than one $\%$ has been obtained on $\Do$ and $\Dp$
lifetimes.

Taking the (naive) average of all measurements the obtained
lifetime ratios are given in Table \ref{tab:liferat}.

The $\Dsp$ lifetime is 20$\%$ larger than the $\Do$ lifetime. A difference
of ${\cal O}(10\%)$ has been anticipated but neither 
its exact value nor its sign
were really predicted \cite{ref:dodslife}. 
The other measured ratios can be used to evaluate 
the relative contributions from P.I. and W.S. mechanisms
\footnote{This has been done very naively considering that the 
respective partial
widths for these processes are independent of the type of the decaying 
$c$-hadron. It has been assumed also that P.I. has the same importance
being constructive or destructive. Possible interferences between different
mechanisms have not been either considered.}.
\begin{eqnarray}
\frac{\Gamma(P.I.)}{\Gamma(spec.)}&\simeq& 0.6 \nonumber \\
\frac{\Gamma(W.S.)}{\Gamma(spec.)}&\simeq&  1.6
\end{eqnarray}

\begin{table*}
\begin{equation}
{\cal A}_{CP}=\frac{|A_f|^2-|\overline{A}_{\bar f}|^2}
{|A_f|^2+|\overline{A}_{\bar f}|^2}=
\frac{2\frac{A_2}{A_1}\sin{(\delta_1^S-\delta_2^S)}
\sin{(\delta_1^W-\delta_2^W)}}
{1+\frac{A_2^2}{A_1^2}+2\frac{A_2}{A_1}\cos{(\delta_1^S-\delta_2^S)}
\cos{(\delta_1^W-\delta_2^W)}}
\label{eq:cqas}
\end{equation}
\end{table*}

\begin{table*}[t]
\begin{center}
\begin{tabular}{|c|c|c|c|c|c|}
\hline
Channel & E791\cite{ref:ecp} & CLEO \cite{ref:cleodgamma}& FOCUS \cite{ref:focuscp}& Average & Theory \cite{ref:cpth}  \\
\hline
$\Do \rightarrow \Km \Kp$ & $-10\pm50$ & $0.5 \pm 23$ & $1 \pm 27$ & $6\pm16$
& $0.1 \pm 0.8$\\
$\Do \rightarrow \pi^- \pi^+$ & $-49 \pm 84$ & $-19.5 \pm 33.3$ & $48 \pm 46$
 & $22 \pm 26$ & $0.02 \pm 0.01$ \\
$\Do \rightarrow {\rm K}^{\circ}_S \pi^{\circ}$ &  & $1\pm13$ & & $1\pm13$ &
 \\
$\Do \rightarrow \pi^{\circ} \pi^{\circ}$ & & $ 1\pm 48$ & & $ 1\pm 48$ & \\
\hline
$\Dp \rightarrow \Km \Kp \pi^+$ & $-14 \pm 29$ & & $6\pm12$ &  $2 \pm 11$ &
$2 \pm 1$ \\
$\Dp \rightarrow \pi^- \pi^+ \pi^+ $ & $-17 \pm 42$ & & & $-17 \pm 42$ &
$-1.2 \pm 0.7$ \\
$\Dp \rightarrow {\rm K}^{\circ}_S \pi^+$ & & & $-5\pm14$ & $-5\pm14$ & \\
$\Dp \rightarrow {\rm K}^{\circ}_S \Kp$ & & & $42\pm52$ & $42\pm52$ & \\
\hline
\end{tabular}
\caption{Present measured values, in $10^{-3}$ units, of CP asymmetries. 
\label{tab:cpcexp}}
\end{center}
\end{table*}

These values indicate that non-spectator effects are similar to the
spectator contribution and not much larger. Thus, in spite of the rather 
low value
of the $c$-quark mass, which is close to 1 GeV/c$^2$, the hierarchy
of charm hadron lifetimes can be understood within the formalism illustrated
by Eq. (\ref{eq:cdecayw}). To go further, absolute semileptonic
branching fractions of $c$-hadrons, especially for the $\Dsp$ 
and $c$-baryons, need to be measured. The same formalism
\cite{ref:volosh} predicts:
\begin{equation}
\Gamma_{sl}(\Omega_c^0) > \Gamma_{sl}(\Xi_c^+) \simeq
\Gamma_{sl}(\Xi_c^0) \geq 2 \Gamma_{sl}(\Lc)
\end{equation}
As an example, the $\Xi_c^+$ and $\Xi_c^0$ are expected to have similar
semileptonic decay widths whereas their lifetimes differ by a factor five.
It can be noted that new lifetime measurements \cite{ref:cloxic} 
\cite{ref:focusxic}
have been obtained 
for the $\Xi_c^+$ which are significantly larger than the previous average.

\subsection{CP violation}
\label{subsec:cp}
As it will clearly appear in Sec. \ref{subsec:oscill}, CP violation
through mixing is expected to be very small for $\Do$ mesons.
The search for effects from direct CP violation seems to be more promising.
CP violation, in those circumstances, requires the contribution 
from two weak and two strong amplitudes in the decay process.
The expression for the CP asymmetry is given in Eq. (\ref{eq:cqas})
in which ${\rm A}_i, ~\delta^S_i~{\rm and}~\delta^W_i,~i=1,2$ are
respectively the moduli, the strong and the weak phases of the two amplitudes.

Penguin graphs, in Cabibbo suppressed decays, are good candidates to provide
the second weak amplitude. Other possibilities involve W.A. graphs in 
$\Dsp$ decays and channels with ${\rm K}^0_S$ \cite{ref:bigisandacp}.
Accurate predictions are difficult because they need to have a control
on strong interaction phases. Typically, effects of ${\cal O}(10^{-3})$
are expected in several channels \cite{ref:cpth}.
Present results of ${\cal A}_{CP}$ are given in 
Table \ref{tab:cpcexp}.

Impressive improvements 
\cite{ref:ecp} \cite{ref:cleodgamma} \cite{ref:focuscp}
have been obtained during the past two years.
On some channels, the sensitivity is close to one $\%$ and systematic
uncertainties do not seem to be the limiting factor. Another
order of magnitude is needed to reach Standard Model expectations
... and more to expect a significant measurement. It can be noted that there 
are also interesting prospects to achieve high sensitivities in
Cabibbo favoured channels (CLEO, FOCUS, $b$-factories) and in 
detailed studies of corresponding Dalitz
plots (even time dependent) for conjugate states.

\subsection{$\ddbar$ oscillations}
\label{subsec:oscill}
\begin{figure}[h]
\epsfxsize180pt
\figurebox{180pt}{180pt}{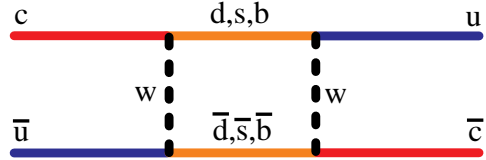}
\caption{An example of a diagram contributing to $\ddbar$ oscillations.}
\label{fig:oscgraph}
\end{figure}

Let's note that this is the only oscillating system involving
a (relatively) heavy up-quark and consequently with $d$-type quarks
circulating inside the loop (see figure \ref{fig:oscgraph}).
This type of diagram generates differences between the masses and the widths
of the mass eigenstates of the $\Do-\Dob$ system. For $b$- and $s$-hadrons,
$\Delta m$ dominates over $\Delta \Gamma$. For charm,
contributions from $d$ and $s$ quarks,
circulating inside the loop, are expected to be dominant
as the values of the CKM matrix elements related to the $b$ quark are 
extremely small. Standard Model expectations for the mass and width differences
between the mass eigenstates of the $\ddbar$ system are extremely small
\cite{ref:ddbarsd}:
\begin{equation}
\frac{\Delta \Gamma}{2\Gamma} |_{sd}\simeq 
~\frac{\Delta m}{\Gamma} |_{sd}\simeq 10^{-4}-10^{-5}
\end{equation}
Corrections from strong interactions can increase these values but not much
apparently \cite{ref:ddbarsd} \cite{ref:bigivolo}:

\begin{equation}
\frac{\Delta \Gamma}{2\Gamma} |_{ld,HQET}\simeq ~\frac{\Delta m}{\Gamma} |_{ld,HQET}\simeq 10^{-3}-10^{-4}
\end{equation}
New physics contributions are expected to enter more probably in 
$\Delta m$ than in $\Delta \Gamma$ whereas non-perturbative QCD contributions,
as for instance violations of parton-hadron duality, can contribute in the
two quantities \cite{ref:bigivolo}. 
From these considerations it results that:
\begin{itemize}
\item experimentally, measuring 
$\frac{\Delta m,~\Delta \Gamma}{\Gamma}\leq 10^{-3}$ is quite challenging,
\item theoretically, having done this measurement 
and found a signal or a limit at this level, is far from trivial, 
implying small effects from parton/hadron duality violation
\cite{ref:bigivolo},
\item it is necessary to measure $\Delta m$ and $\Delta \Gamma$ in a separate
way.
\end{itemize}
For a recent update on this subject see also \cite{ref:petrov}.

\begin{figure}[h]
\epsfxsize180pt
\figurebox{180pt}{180pt}{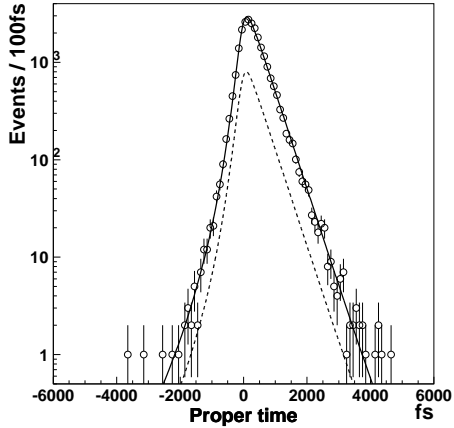}
\caption{Proper-time distribution and lifetime fit result
for the $\Do \rightarrow \Km \Kp$ decay mode. 
Events have been selected in the $\Do$ mass signal region.
The dotted line shows the background contribution in the fit.
Figure from \protect\cite{ref:belledgamma}.}
\label{fig:kkli}
\end{figure}

Recent results have been obtained by BELLE \cite{ref:belledgamma}
and CLEO \cite{ref:cleodgamma} on $\Delta \Gamma$.
These analyses consist in measuring the lifetime for events corresponding
to different combinations of mass eigenstates
\footnote{It has been assumed, in the following, that mass and CP eigenstates are the same.}. 
For instance, $\Kp \Km$
or $\pi^+ \pi^-$ decay channels correspond to a CP (or mass) eigenstate
whereas $\Km \pi^+$ contains the same fraction of the two mass eigenstates.
The measured lifetimes (see figure \ref{fig:kkli})
can be related to the decay width 
difference between the two mass eigenstates of the $\ddbar$
system:
\begin{equation}
\frac{\Delta \Gamma}{2 \Gamma}=\frac{\tau(\Km \pi^+)}{\tau(\Km \Kp)}-1
\end{equation}

\begin{table*}[t]
\begin{equation}
R_{WS}(t)=\left |\frac{<\Km \pi^+|\Dob(t)>}
{<\Km \pi^+|\Do(t)>} \right |^2= R_{DCS} +
\sqrt{R_{DCS}}~ y^{\prime}
\left ( \frac{t}{\tau(\Do)}\right )
+ \frac{x^{\prime 2}+y^{\prime 2}}{4}
\left ( \frac{t}{\tau(\Do)}\right )^2 
\label{eq:kpit}
\end{equation}
\end{table*}

\begin{figure}[h]
\vspace*{-0.5cm}
\epsfxsize180pt
\figurebox{180pt}{180pt}{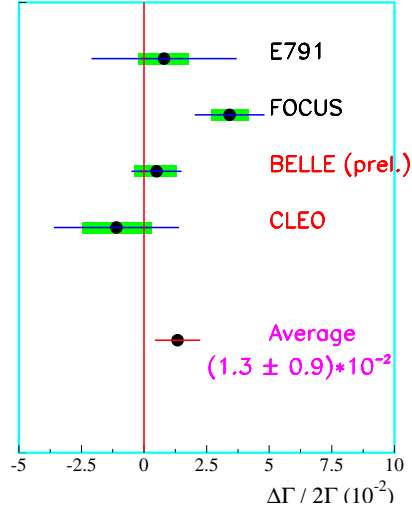}
\caption{Present measurements of $\frac{\Delta \Gamma}{2 \Gamma}$
and global average.}
\label{fig:alldgdo}
\end{figure}

The new  measurements have not confirmed the positive effect seen
by FOCUS \cite{ref:focusdgamma}
as shown in figure \ref{fig:alldgdo}. Averaging all results gives
a value compatible with zero and it must be noted that the sensitivity
has already reached 1$\%$ and does not seem to be limited by systematics.
There are thus good prospects for improvements at $b$-factories.

Other types of measurements are sensitive both to $\Delta m$ and 
$\Delta \Gamma$. They consist in comparing the time evolution
of Cabibbo favoured (C.F.) and doubly Cabibbo suppressed (D.C.S.)
decays. An example is given in Eq. (\ref{eq:kpit})
for the $\Do \rightarrow \Km \pi^+$ channel.

This expression has three components:
$R_{DCS}$ corresponds to the D.C.S. decay rate normalized to the C.F. rate,
the term with the quadratic $t$ dependence is due to oscillations whereas
the interference between the two processes has a linear time dependence.
This expression depends on four parameters: $R_{DCS}$, 
$x=\frac{\Delta m}{\Gamma}$, $y=\frac{\Delta \Gamma}{2\Gamma}$ and $\delta$.
The last quantity corresponds to a possible strong phase difference
for $\Do$ decaying into $\Km \pi^+$ or  $\Kp \pi^-$ respectively.
As a result, in Eq. (\ref{eq:kpit}), the parameters $x^{\prime}$ 
and $y^{\prime}$, which 
are entering, are related to $x$ and $y$ by a rotation of angle $\delta$:
\begin{equation}
x^{\prime}= x \cos{(\delta)} + y \sin{(\delta)}
\end{equation}
and
\begin{equation}
y^{\prime}= -x \sin{(\delta)} + y \cos{(\delta)}
\end{equation}

\begin{figure}[!h]
\epsfxsize180pt
\figurebox{180pt}{180pt}{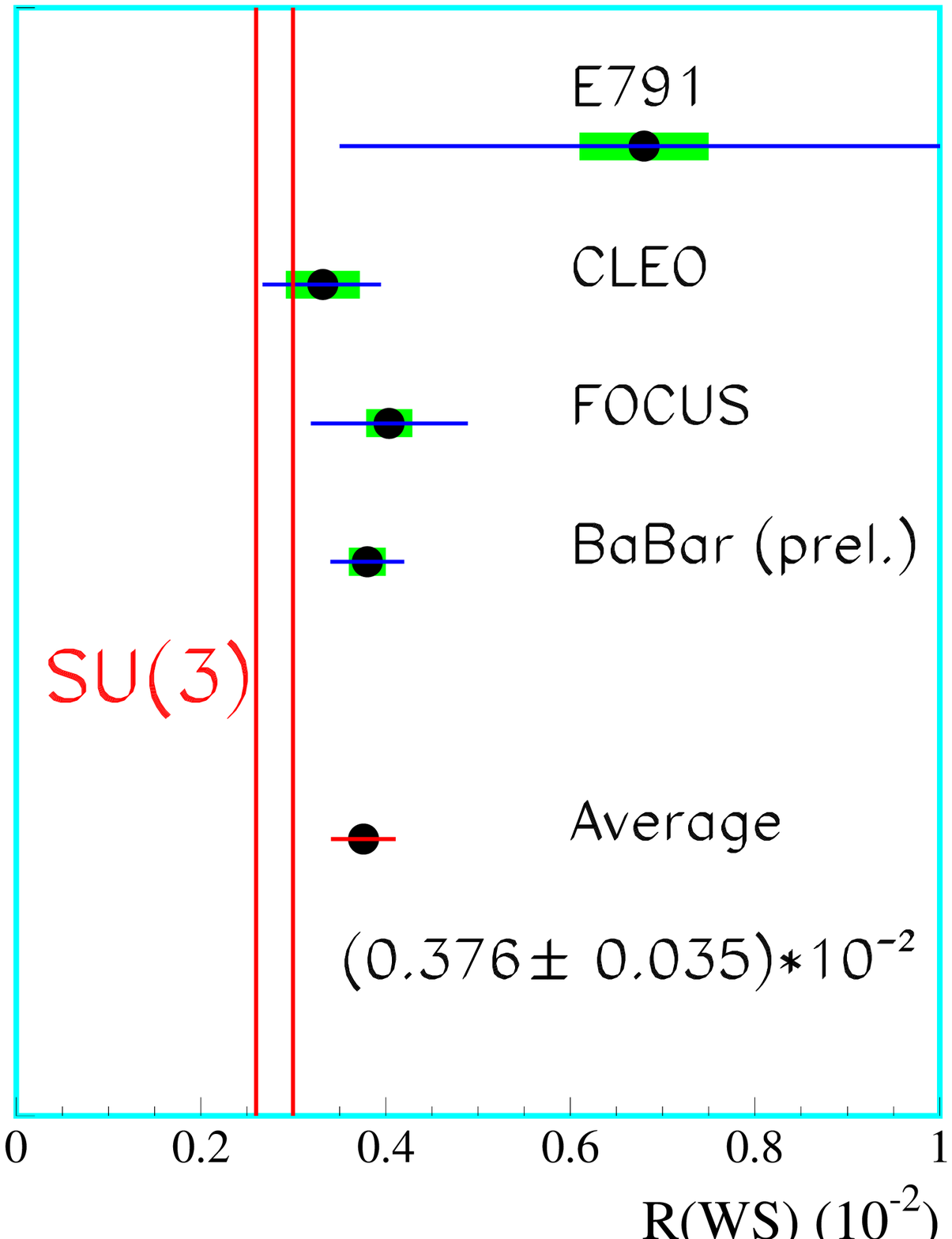}
\caption{Measurements and global average of wrong sign 
relative to right sign decay rates of $\Do$ mesons
into K$\pi$. Vertical lines correspond to the naive expectation:
$R_{WS}=\tan^4{(\theta_c)}$.}
\label{fig:wsd}
\epsfxsize180pt
\figurebox{180pt}{180pt}{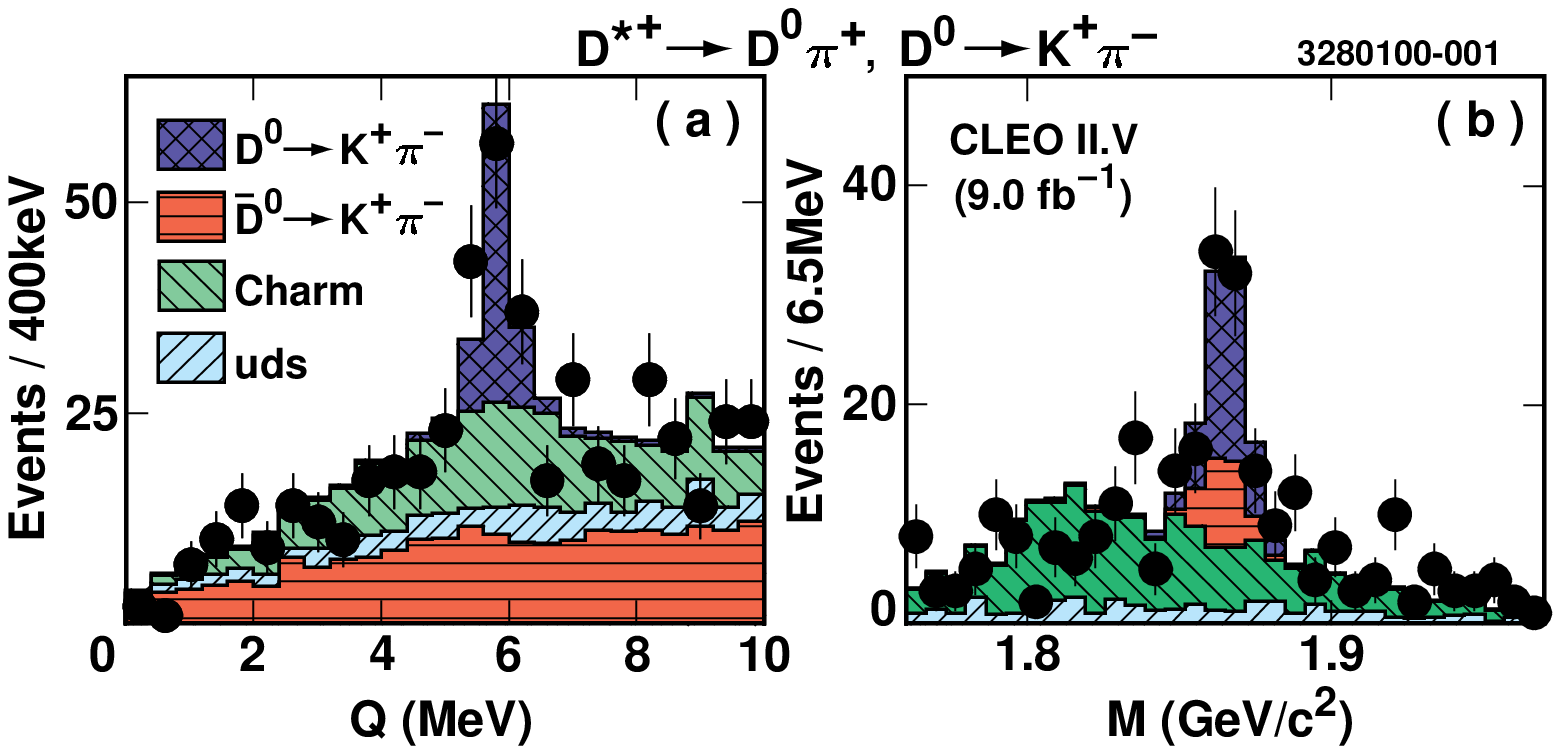}
\caption{Signal for the W.S. process $\Do \rightarrow \Kp \pi^-$.
The data are the full circles with error bars, the projection of the fit
for the signal is crosshatched, and the projections of the fit for the
backgrounds from charm and light quark production are singly hatched. 
For part (a), M is within 2$\sigma$ of the C.F. value, and for (b), 
$Q=m({\rm K} \pi \pi)-m({\rm K} \pi)-m(\pi)$ is within
2$\sigma$ of the C.F. value. Figure from \protect\cite{ref:cleows}.}
\label{fig:cleows}
\end{figure}

Impressive progress has been made also in the determination of the integrated
rate of wrong sign 
\footnote{The D.C.S. and W.S. rates have to be distinguished.
The W.S. rate corresponds to the integral of W.S. events time 
distribution. For experiments without any bias or limits on the 
reconstructed $\Do$ meson proper time:
$R_{WS}=R_{DCS}+\sqrt{R_{DCS}}~y^{\prime}+\frac{x^{\prime 2}+y^{\prime 2}}{2}$.}
(W.S.) $\Do$ hadron decays as illustrated in figure \ref{fig:wsd}.
New results have been obtained by CLEO \cite{ref:cleows}
 (figure \ref{fig:cleows})
and BaBar  \cite{ref:babarws} collaborations.
Combining the different measurements,
 $R_{WS}$ is obtained with 10$\%$
accuracy. There are also new measurements from CLEO \cite{ref:cleoautrews}
on other W.S. $\Do$ decay channels:
\begin{equation}
R_{WS}(\Km \pi^+ \pi^0)=(0.43^{+0.11}_{-0.10}\pm0.07)\times 10^{-2}
\end{equation}
\begin{equation}
R_{WS}(\Km \pi^+ \pi^+ \pi^-)=(0.41^{+0.12}_{-0.11}\pm0.04)\times 10^{-2}
\end{equation}

\begin{figure}[h]
\epsfxsize180pt
\figurebox{180pt}{180pt}{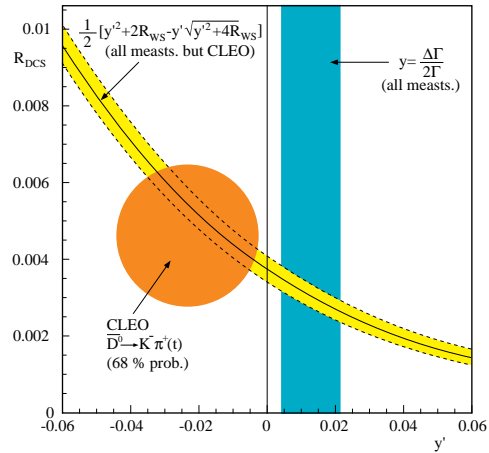}
\caption{Summary of present measurements in the plane ($y^{\prime},~R_{DCS}$).
The curved region corresponds to the constraint given by the measured
rate of W.S. events, considering that the last term
in Eq. (\ref{eq:kpit}) has a negligible contribution.
The circle is the result from CLEO \protect\cite{ref:cleows} obtained by
fitting their measured time distribution (and neglecting the correlation
between the two fitted values). 
The vertical band corresponds 
to the averaged value of $y$ given in figure \ref{fig:alldgdo}.}
\label{fig:allrwsdcs}
\end{figure}

As long as $x$ and $y$ are small, as compared with
$\sqrt{R_{DCS}}\simeq \theta_c^2 \simeq 0.05$, the linear term dominates
over the last one,
in Eq. (\ref{eq:kpit}). Integrating over the time variable,
results can be expressed in the $(y^{\prime},~R_{DCS})$ plane,
see figure \ref{fig:allrwsdcs}. 
In this figure, 
the CLEO measurement, obtained by fitting 
$R_{DCS},~x^{\prime}~{\rm and}~y^{\prime}$ on their registered 
time distribution
 has been given along
with the constraint obtained by combining wrong-sign rate measurements
including all results but CLEO. Within 
one standard deviation (or so) these results are compatible and 
also compatible with zero. If the phase $\delta$ is not too close to zero
(30$^{\circ}$ or more) and if $x$ is not much smaller than $y$,
it can be expected that a comparison between the values of $y$ and 
$y^{\prime}$ gives access to $x$. If not, it seems extremely difficult
to be sensitive 
to values of $x$ below 1$\%$ because of the quadratic dependence
of the last term in Eq. (\ref{eq:kpit}). 
Extracting $x$ from measurements of $y$ an $y^{\prime}$
requires a determination
of $\delta$ which seems to be possible at a $c$-factory
\cite{ref:rosnerphase}.

In the coming years it can be expected that analyses
of W.S. events measured time distributions allow one to reach
an accuracy on $y^{\prime}$ of a few $10^{-3}$.

\begin{table*}[t]
\begin{center}
\begin{tabular}{|c|c|c|c|}
\hline
  Sample & Nominal & Kinematic cuts & Tracking quality cuts \\
\hline
$\#$ events & 11496& 3284 & 368\\
bad measts. & $\simeq 5\%$ & none & none \\
 fitted width (keV)& $96.2 \pm 4.0 $ & $103.8 \pm 5.9$ & $104 \pm 20$\\
systematics (keV) & $\pm22$& $\pm20$ &$\pm22$ \\
\hline
\end{tabular}
\caption{Summary of the data samples, simulation biases, and fit results.}
\label{tab:selectgds}
\end{center}
\begin{equation}
\Gamma(\Dstarp)=\frac{g^2}{24\pi m^2_{D^*}}p^3_{\pi^+}
+\frac{g^2}{48\pi m^2_{D^*}}p^3_{\pi^{\circ}}
+\Gamma(\Dp \gamma)
\label{eq:gdstar}
\end{equation}
\end{table*}

\subsection{$\Dstarp$ width measurement}
\label{subsec:dstarw}
The measurement of the $\Dstarp$ width is challenging because it needs
a spectrometer of high resolution and a detailed understanding of its
properties. A difficult point comes from the absence of a monitoring
channel. In CLEO \cite{ref:dstarwidth}, this has been accomplished 
by performing a detailed comparison 
between the characteristics of reconstructed and simulated charged 
particle trajectories. The studied decay chain is:
$\Dstarp\rightarrow \Do \pi^+,
~\Do\rightarrow \Km \pi^+$.

\begin{figure}[h!]
\epsfxsize180pt
\figurebox{180pt}{180pt}{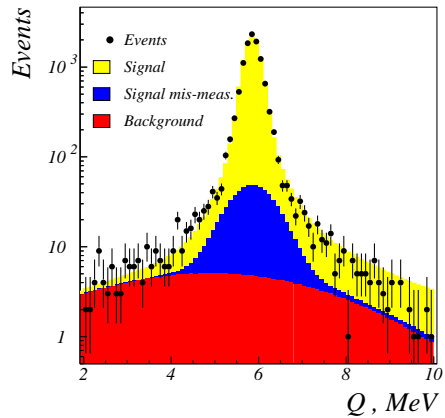}
\caption{Fit to nominal data sample. The different contributions to the fit 
are shown. Figure from \protect\cite{ref:dstarwidth}.}
\label{fig:selectgds}
\end{figure}

In figure \ref{fig:selectgds}, distributions of the variable 
$Q=m(\Km \pi^+ \pi^+)-m(\Km \pi^+)-m(\pi^+)$, for real and simulated
events are compared and the value of the intrinsic $\Dstarp$ width
is extracted. Typical experimental resolutions on $Q$ are of the order
of 150 KeV. Three samples of events, selected with criteria, providing 
different sensitivities 
of the measured $Q$ value, to detector properties, due to decay characteristics
of the events or to the quality of charged particle reconstruction,
have been analysed. They give similar results as indicated in Table 
\ref{tab:selectgds}:

\begin{equation}
\Gamma(\Dstarp)=(96\pm4\pm22)~{\rm KeV}
\end{equation}
 An example of a fitted distribution is given in figure
\ref{fig:selectgds}.

This measurement is important because it allows one to determine the value of
the amplitude for soft pion emission in charm decays:
\begin{eqnarray}
g=g_{D^*D^+\pi^{\circ}}&=&\frac{g_{D^*D^{\circ}\pi^+}}{\sqrt{2}}~{\rm (isospin)} \nonumber \\
&=&17.9\pm1.9
\end{eqnarray}

The expression relating the $\Dstarp$ total width and $g$ 
 which can be 
found, for instance in \cite{ref:gdstartog},
is given in Eq. (\ref{eq:gdstar}).

The amplitude for soft pion emission enters in chiral evaluations and also, 
as an example, in single pole form factor parametrizations of the semileptonic
decay ${\rm D}\rightarrow \pi \ell^{+} \nu_{\ell}$:
\begin{equation}
f_{D\pi}^+(q^2)=\frac{f_{D*}~g}{2 m_{D^*}}
\left(\frac{1}{1-\frac{q^2}{m_{D^*}^2}}\right)
\label{eq:slff}
\end{equation}
which can be used when $q^2 \rightarrow m_D^2$ so that the emitted pion 
is soft.
Now the problem comes when the actual result from CLEO is compared with
expectations \cite{ref:yaou}. 
Apart for quark models which are in reasonable agreement
with the experimental result, but for which it
is difficult to attribute an uncertainty, 
other more QCD-based approaches are failing
by a large amount. Light cone sum rules allow one to evaluate the product
$f_D f_{D^*} g$ and, separately, $f_D$ and $f_{D^*}$. They predict
\cite{ref:lcsr} a rather low value:
\begin{equation}
\Gamma(\Dstarp)=(35\pm20)~{\rm KeV}
\end{equation}
In lattice QCD there is not yet a direct evaluation of $g$ for charm particles.
Using their predicted value for the semileptonic form factor
at zero recoil
($f_{D\pi}^+(0)=0.64\pm0.05^{+0.}_{-0.07}$ \cite{ref:abamar}) 
and Eq. (\ref{eq:slff})
in which the measured value for $g$ is used, a value for the 
$\Dstarp$ decay constant is obtained: $f_{D^*}=140$ MeV.
This value is rather low as compared with expectations of the order
of $270$ MeV\footnote{It is expected that $f_{D^*}>f_{D}\sim240~ MeV$.}.
But it must be noted that the result obtained from lattice QCD for
$f_{D\pi}^+(0)$ is in rather good agreement with the value deduced
from QCD light-cone sum rules and with 
the experimental measurement of the semileptonic branching fraction
${\rm BR}(\Do \rightarrow \pi^- e^+ \nu_e)=(3.7 \pm 0.6)\times 10^{-3}$
\cite{ref:pdg00}
which, using the parametrization of Eq. (\ref{eq:slff}), 
corresponds to $f_{D\pi}^+(0)$ = $0.74 \pm 0.03$.
It has to be emphasized that Eq. (\ref{eq:slff}) is expected
to be valid at high $q^2$ whereas the differential decay rate is
negligible in this region. Consequently, when comparing the total
decay rate with expectations from lattice QCD or QCD sum rules, the $q^2$ 
dependence of the form factor is also relevant. Present discrepancies
may point to a more complicated $q^2$ variation than given in
Eq. (\ref{eq:slff}).
It seems rather important that theorists examine in some detail the present 
situation.

From the experimental side, it is important also that this measurement be 
repeated at $b$-factories, which seem to have similar capabilities as CLEO,
and possibly reduce the present systematic uncertainty of $\pm22$ KeV.

\subsection{Absolute branching fraction measurements}
The present situation \cite{ref:pdg00}, summarized
in Table \ref{tab:brd}, needs to be improved a lot.
\begin{table}[h]
\begin{tabular}{|c|c|c|c|}
\hline
  & Exclusive & $D \rightarrow \ell^+ X$ & $ D \rightarrow \Km X$\\
\hline
$\Do$ & $2.3\%$   &  $4.5\%$ & $8\%$ \\
$\Dp$ & $6.7\%$   &  $11\%$ & $12\%$ \\
$\Ds$ & $25\%$   &  $75\%$ & $100\%$ \\
$\Lc$ & $26\%$   &  ? & ? \\
$\Xi_c^{\circ}$ & ?    &  ? & ? \\
$\Xi_c^+$ & ?    &  ? & ? \\
$\Omega_c^+$ & ?   &  ? & ? \\
\hline
\end{tabular}
\caption{Present relative accuracies on the best measured
$c$-hadrons exclusive
decay channels and on two inclusive processes. 
The sign ``?'' indicates that no information is available.}
\label{tab:brd}
\end{table}

For exclusive final states, only $\Do$ decay channels are measured
with some accuracy even if the few permil level, reached in $\tau$
decays, is far from being accessed. The determination of inclusive 
properties is really poor. The reason for this situation is very
clear. At all $e^+-e^-$ colliders $\tau$ leptons are produced in pairs
without any other accompanying particle and thus, absolute branching
fraction measurements are a priori possible at all these machines
\footnote{Background characteristics are not completely negligible
as, for instance, the ALEPH \cite{ref:alephpp} experiment at LEP
has obtained similar precisions as CLEO \cite{ref:cleotaupipio}
on the $\tau^- \rightarrow \pi^- \pi^0 \nu_{\tau}$ channel with
ten times less statistics.}. For charm,
the only place were $c$-hadrons are produced in pairs of conjugate states,
is in the threshold region. As the present situation is extremely poor
and as large statistics of $c$ and $b$-decays are being registered
at CLEO and $b$-factories, dedicated tagging procedures can
be developed and improved results are expected. As an example, the recent 
measurement, by CLEO \cite{ref:cleolcabs}, 
of the $\Lc$ branching fraction into $p \Km \pi^+$
can be given. It is difficult to imagine that such approaches can reach
accuracies better than several percent.

The accurate determination of $\Dp$ and $\Ds$ decay constants
and of the $q^2$ dependence of form factors in semileptonic decays,
which are important measurements to validate lattice QCD calculations
can be made only at a Charm factory. This is true also for the determination
of inclusive properties of $c$-hadrons for which,
as an example, the most precise 
results on the $\Dp$ are still coming from MARKIII \cite{ref:markd}.

\subsection{D decay constant measurements}
\begin{figure}
\epsfxsize180pt
\figurebox{180pt}{180pt}{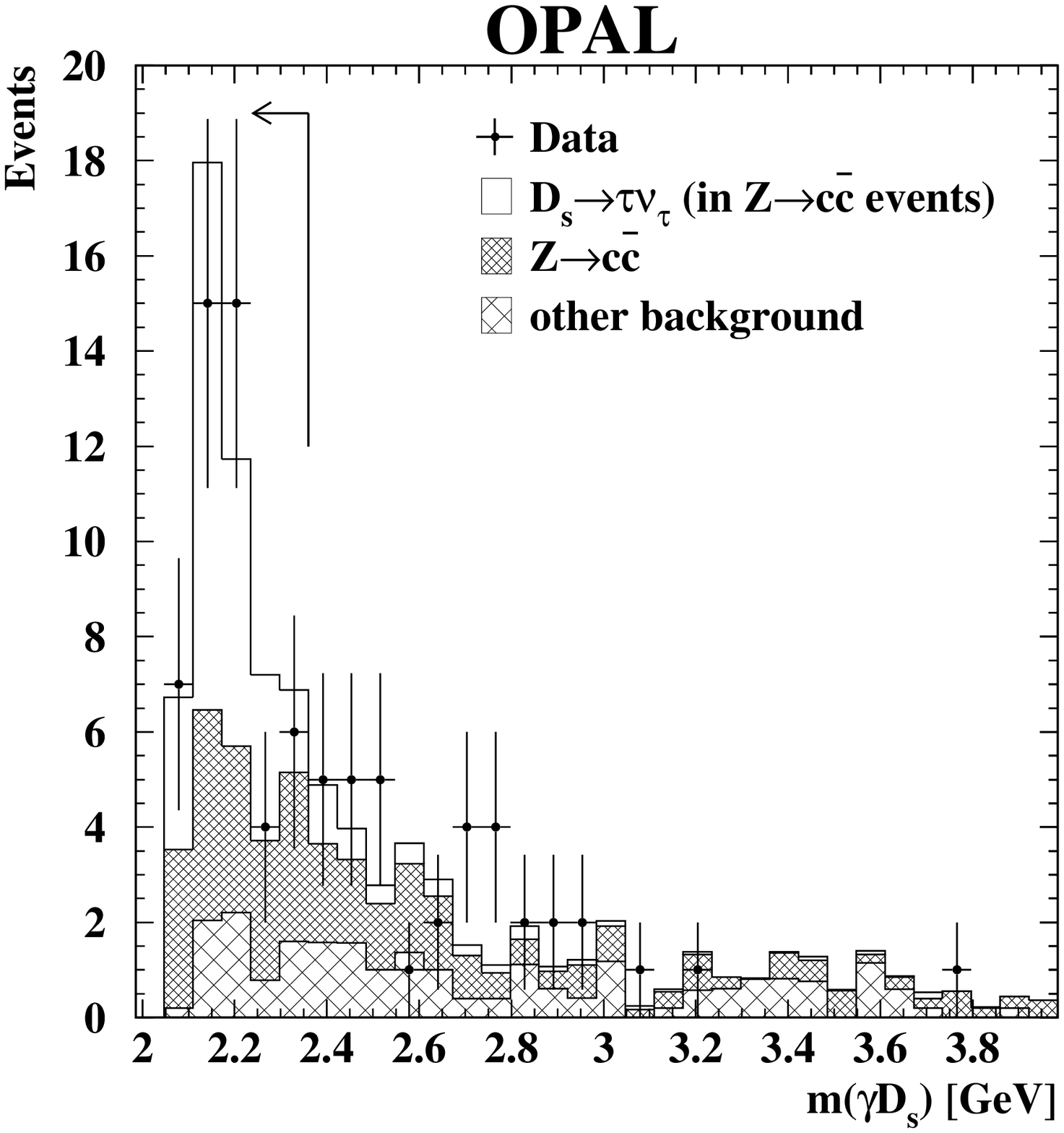}
\caption{Invariant mass $m(\gamma\Ds)$ of the photon and the $\Ds$
candidate for the events satisfying all selection criteria.
The contributions to the Monte Carlo distribution from the signal and from
the different sources of background are shown separately. The signal region
is indicated by an arrow. Figure from \protect\cite{ref:opalfds}.}
\label{fig:opalfd}

\epsfxsize180pt
\figurebox{180pt}{180pt}{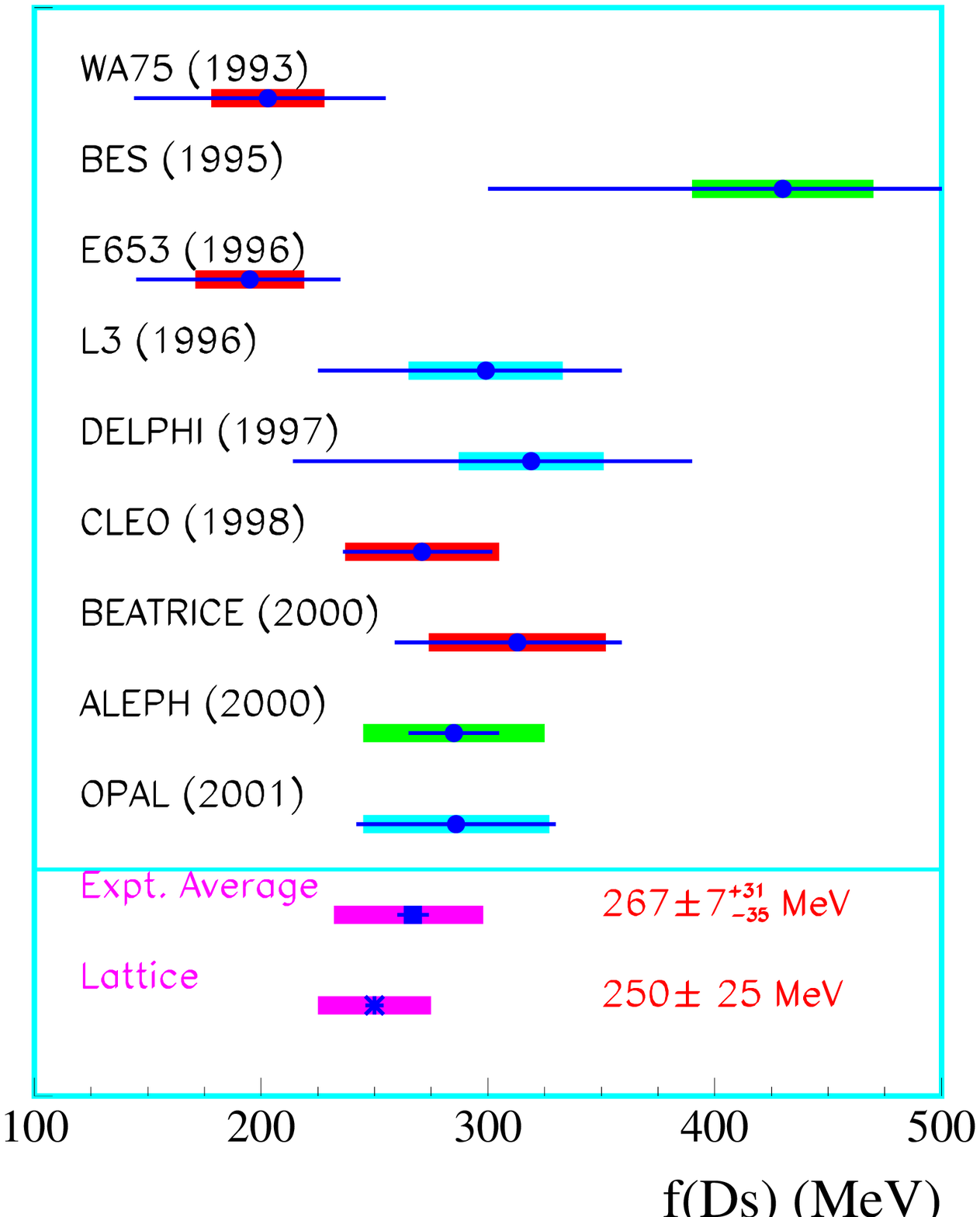}
\caption{Summary of experimental results on the $\Ds$ decay constant.
Measurements of $\Ds \rightarrow \tau^+ \nu_{\tau}$ and
$\Ds \rightarrow \mu^+ \nu_{\mu}$ decay rates have been combined.}
\label{fig:allfd}
\end{figure}

A recent result has been obtained by OPAL \cite{ref:opalfds}
(see figure \ref{fig:opalfd}) and ALEPH \cite{ref:alephfds}
has also produced a new report.
All measurements have been summarized in figure \ref{fig:allfd}. The leptonic branching fraction of charged D mesons is given in Eq. 
(\ref{eq:brfd}) which is similar to the corresponding
expression for pions. Because of helicity conservation, the decay rate
is proportional to the square of the charged lepton mass. It follows that decays into $\tau$ leptons are favoured whereas those
into electrons can be neglected. Typical values for expected branching
fractions are:
\begin{table*}[t]
\begin{equation}
{\rm BR}({\rm D}_{s,d} \rightarrow \ell^+ \nu_{\ell}) = \frac{{\rm G_F}^2}{8 \pi}
\tau({\rm D}_{s,d})            f_{D_{s,d}}^2  \left | {\rm V}_{c(s,d)} \right |^2
m_{D_{s,d}} m_{\ell}^2 \left ( 1-\frac{m_{\ell}^2}{m_{D_{s,d}}^2} \right )^2
\label{eq:brfd}
\end{equation}
\end{table*}

\begin{eqnarray}
{\rm BR}(\Ds \rightarrow \tau^+ \nu_{\tau})& =& 5.7\%,\nonumber \\
{\rm BR}(\Ds \rightarrow \mu^+ \nu_{\mu})&=& 0.6\%,\nonumber \\
{\rm BR}(\Dp \rightarrow \tau^+ \nu_{\tau}) &=& 0.1\%,\nonumber \\
{\rm BR}(\Dp \rightarrow \mu^+ \nu_{\mu})&=& 0.04\%
\end{eqnarray}
taking $f_{D_d}=220$ MeV and $f_{D_s}=260$ MeV.
The OPAL analysis  \cite{ref:opalfds} is using neural networks to reduce 
the contamination
from $\Zz \rightarrow b\overline{b}$ events and from other $c$-hadron decays
in $\Zz \rightarrow c\overline{c}$ events. Evidence for the signal
is obtained using the cascade decay $\Ds^{*} \rightarrow \Ds \gamma$
(see figure \ref{fig:opalfd}).

The average value for $f_{D_s}$, taking into account correlated systematic 
uncertainties which are dominated by the error on the 
$\Ds \rightarrow \phi \pi^+$ branching fraction, is equal 
\footnote{A rather similar value: $f_{D_s}=(264 \pm 15 \pm 33)$ MeV
can be found in \cite{ref:autrefds}.}to:
\begin{equation}
f_{D_s}=(267 \pm 7 ^{+31}_{-35})~{\rm MeV}
\end{equation}
and agrees with lattice QCD evaluations which correspond to
$f_{D_s}(lattice)=(250 \pm 25)$ MeV \cite{ref:ciuchifds}.
Such a comparison is expected to be a stringent test for lattice QCD 
as measurements of D meson decay constants can reach
a percent accuracy at a Charm facility. Obtaining values of B meson decay
constants from extrapolations of measured D decay constants could be 
a valuable approach \cite{ref:prach} as it is impossible, 
in practice, to measure directly
$f_B$.

\section{Conclusions}

As you have noted $\tau$ and charm physics are on very different grounds.
They are extremely nice laboratories to study QCD in its perturbative and 
non-perturbative aspects. These properties have been already well 
exploited in $\tau$ decays whereas, for charm, a lot remains to be done.

The two aspects are complementary. In $\tau$ decays the production
of hadrons is clean and simple, through the charged weak current. This
allows one to control non-perturbative contributions and to access fondamental 
parameters of the theory such as $\alpha_s$ and quark masses. In charm decays, 
the hadronic current is coupled to a heavy quark (charm). Simple
situations met in leptonic and semileptonic decays allow one to control 
lattice QCD evaluations of heavy meson decay constants and form factors.
Accurate measurements in these domains need the use of a charm factory.
This will open a basic understanding of non-perturbative QCD
for the benefit of, for instance, the extraction of CKM matrix 
element values from measurements of $b$-hadron decays or oscillations.

\vspace*{-0.2cm}
\section*{Acknowledgments}
I would like to thank all members of the different Collaborations
who sent me new results, in time for the Conference. I apologize
for not having been able to give a complete account of all their
achievements. I thank A. Le Yaouanc for having explained to me
the different problems raised by the CLEO measurement of the $\Dstar$
width and R. R\"uckl for his comments on this subject. I have benefitted
also of useful comments from I.I. Bigi,
especially on charm lifetime measurements and from M. Davier and A. Pich
on the $\tau$ aspect of this presentation.
I thank the organizers of the Conference for their invitation and support.
Special thanks to Mario Antonelli for his help during the Conference.
Merci \'egalement \`a C. Bourge pour son aide technique 
durant l'\'elaboration de ce document.

\end{document}